\documentclass[oneside,english]{amsart}
\usepackage[T1]{fontenc}
\usepackage[latin9]{inputenc}
\usepackage[active]{srcltx}
\usepackage{bm}
\usepackage{amsthm}
\usepackage{amstext}
\usepackage{graphicx}
\usepackage{esint}

\makeatletter
\numberwithin{equation}{section}
\numberwithin{figure}{section}
  \theoremstyle{remark}
  \newtheorem*{rem*}{\protect\remarkname}

\pdfoutput=1
\usepackage{babel}

\makeatother

\usepackage{babel}
  \providecommand{\remarkname}{Remark}

\begin{document}

\title{A fick-Jacobs equation for channels over 3D curves }

\author{Carlos Valero Valdes\\
 Departamento de Matematicas Aplicadas y Sistemas\\
 Universidad Autonoma Metropolitana-Cuajimalpa\\
 México, D.F 01120, México\\
 \\
 Rafael Herrera Guzman\\
 Centro de Investigacion en Matematicas (CIMAT)\\
 Guanajuato, Gto\\
 México.}

\date{04 November 2014}

\thanks{Partially supported by PROMEP grantUAM-PTC-379 and CONACyT grant
135106. }
\begin{abstract}
The purpose of this paper is to provide a new formula for the effective
diffusion coefficient of a generalized Fick-Jacobs equation for narrow
3-dimensional channels. The generalized Fick-Jacobs equation is obtained
by projecting the 3-dimensional diffusion equation along the normal
directions of a curve in three dimensional space that roughly resembles
the narrow channel. The projection (or dimensional reduction) is achieved
by integrating the diffusion equation along the cross sections of
the channel contained in the planes orthogonal to the curve. We show
that the resulting formula for the associated effective diffusion
coefficient can be expressed in terms of the geometric moments of
the channel's cross sections and the curve's curvature. We show the
effect that a rotating cross section with offset has on the effective
diffusion coefficient. 
\end{abstract}

\maketitle
\global\long\def\CC{\mathbb{C}}

\global\long\def\RR{\mathbb{R}}

\global\long\def\map{\rightarrow}

\global\long\def\mult{\mathcal{M}}

\global\long\def\EE{\mathcal{E}}

\global\long\def\OO{\mathcal{O}}

\global\long\def\FH{\mathcal{F}}

\global\long\def\CH{\mathcal{H}}

\global\long\def\VV{\mathcal{V}}

\global\long\def\PP{\mathcal{P}}

\global\long\def\CS{\mathcal{C}}

\global\long\def\tangent{T}

\global\long\def\cotangent{\tangent^{*}}

\global\long\def\conormal{\mathcal{C}}

\global\long\def\SS{\mathcal{S}}

\global\long\def\KK{\mathcal{K}}

\global\long\def\NN{\mathcal{N}}

\global\long\def\sym#1{\hbox{S}^{2}#1}

\global\long\def\symz#1{\hbox{S}_{0}^{2}#1}

\global\long\def\proj#1{\hbox{P}#1}

\global\long\def\SO{\hbox{SO}}

\global\long\def\GL{\hbox{GL}}

\global\long\def\U{\hbox{U}}

\global\long\def\tr{\hbox{tr}}

\global\long\def\ZZ{\mathbb{Z}}

\global\long\def\der#1#2{\frac{\partial#1}{\partial#2}}

\global\long\def\dder#1#2{\frac{\partial^{2}#1}{\partial#2^{2}}}

\global\long\def\covder{\hbox{D}}

\global\long\def\diff{d}

\global\long\def\dero#1{\frac{\partial}{\partial#1}}

\global\long\def\ind{\hbox{\, ind}}

\global\long\def\deg{\hbox{deg}}

\global\long\def\smb{S}

\global\long\def\DD{\mathcal{D}}

\global\long\def\QQ{\mathcal{Q}}

\global\long\def\RRR{\mathcal{R}}

\global\long\def\dd#1#2{\frac{d^{2}#1}{d#2^{2}}}

\global\long\def\d#1#2{\frac{d#1}{d#2}}

\global\long\def\and{\hbox{\,\, and\,\,\,}}

\global\long\def\where{\hbox{\,\, where\,\,\,}}

\global\long\def\KK{\mathcal{K}}

\global\long\def\II{\mathcal{I}}

\global\long\def\JJ{\mathcal{J}}

\global\long\def\Re{\hbox{Re}}

\global\long\def\Im{\hbox{Im}}

\section{Introduction}

Understanding spatially constrained diffusion in quasi-one dimensional
systems is of fundamental importance in various sciences, such as
biology (e.g. channels in biological systems), chemistry (e.g. pores
in zeolites) and nano-technology (e.g. carbon nano-tubes). However,
solving the diffusion equation in arbitrary channels is a very difficult
task. One way to tackle it, which we follow in this paper, consists
in reducing the degrees of freedom of the problem by considering only
the main direction of transport.

The study of diffusion in (nearly) planar narrow channels has been
undertaken and developed by several authors \cite{kn:kp-diffusion-projection,kn:bradley,kn:di-projection-diffusion}
following the approach of reducing the dimensionality of the problem
to one dimension. They have provided formulas for estimates of the
effective diffusion coefficient by \textquotedbl{}projecting\textquotedbl{}
the two dimensional diffusion equation onto a straight line. More
recently (see \cite{kn:projdiff}), we have generalized this work
by projecting the 2-dimensional diffusion onto an arbitrary curve
on the plane, thus providing estimates of the effective diffusion
coefficient involving the geometrical information of the curve (i.e.
its curvature).

In all the work mentioned above we can distinguish two cases: the
infinite transversal diffusion rate case and the finite transversal
diffusion rate case. In the former, it is assumed that the concentration
distribution stabilizes instantly in the transversal directions of
the channel and, in the latter, the finite time of transversal stabilization
is taken into account. In mathematical terms this cases can be characterized
as follows. In the first case the effective diffusion coefficient
only involves $0$-th order geometrical quantities of the channel
(such as width). In the second case this coefficient involves higher
order geometrical information, such as that arising from the tangential
and curvature information (i.e. higher derivatives) of the channel's
surface wall(s).

On the other hand, the diffusion process in 3-dimensional (non-planar)
channels presents more complications and remains a difficult problem
to tackle. Some attempts have been carried out by Ogawa \cite{kn:ogawa},
Kalinay \& Percus \cite{kn:kp-diffusion-projection}, Antipov et al
\cite{kn:antipov}. Owaga derived a formula for the effective diffusion
coefficient for channels in 3-dimensional space over a central curve
with constant rectangular cross section, and showed that the curvature
of the central curve plays a fundamental role. Kalinay and Percus
studied the case of a hyporboloidal cone. Antipov et al. studied the
case of a periodically expanding and contracting straight channel.

The main motivation for using arbitrary curves in the dimensionality
reduction technique is the following: by choosing a curve that \textquotedbl{}follows\textquotedbl{}
the channel's geometry as closely as possible, one is able to provide
better estimates of the effective diffusion coefficient. If fact,
we have shown in \cite{kn:projdiff} that for two dimensional channels
which are symmetric and of constant width, the formulas for the effective
diffusion coefficient coincide in the finite and infinite transversal
diffusion rate cases. In \cite{kn:ogawa} Ogawa proved the same result
for 3-dimensional channels having constant rectangular cross section.

Thus, the purpose of this paper is to derive a new formula for the
effective diffusion coefficient (in the infinite transversal diffusion
rate case) for 3-dimensional channels defined around a central curve
in 3-dimensional space whose orthogonal cross section is not necessarily
constant. We derive a formula for the effective diffusion coefficient
with dependence on the curvature of the base curve, and the geometric
and \textquotedbl{}statistical\textquotedbl{} properties of the cross
section (i.e. its geometric moments). In particular, we derive explicit
formulas relating the effective diffusion coefficient to the average
widths and average rotation of the cross section of the channel with
respect to the Frenet-Serret moving frame of the curve.

The outline of our article is as follows:
\begin{itemize}
\item In section \ref{effective-continuity}, we will show how the three
dimensional continuity equation on a channel can be reduced to a one
dimensional continuity equation. This last equation, which we will
call the effective continuity equation, will serve as the basis for
what follows in the rest of the article.
\item In section \ref{generalized-fick-jacobs}, we will derive a generalized
Fick-Jacobs equation and a new formula for the effective diffusion
coefficient $\DD$ corresponding to the infinite transversal diffusion
rate case (see formula (\ref{eq:effective-diffusion-coefficient})).
The standard Fick-Jacobs equation corresponds to the case when the
base curve has zero curvature (i.e. it is a straight line). We will
use standard tools of differential geometry of 3-dimensional curves
to write down the formula for $\DD$.
\item In section \ref{applications}, we study channels with gyrating cross
section and deduce Ogawa's formula \cite{kn:ogawa} as a particular
case.
\item We finish with conclusions in section \ref{conclusions}, a brief
review of the necessary differential geometric material in Appendix
1, and in Appendix 2 we provide the details of the computations used
to obtain some explicit formulas for the effective diffusion coefficient
functions. 
\end{itemize}

\section{The effective continuity equation on a 3-dimensional Region}

\label{effective-continuity}

We are interested in describing a transport process on a channel-like
region $\Omega$ in 3-dimensional space (see Figure \ref{fig:OmegaAndSAtU}).

\subsection*{The continuity equation}

\begin{figure}
\includegraphics[scale=0.5]{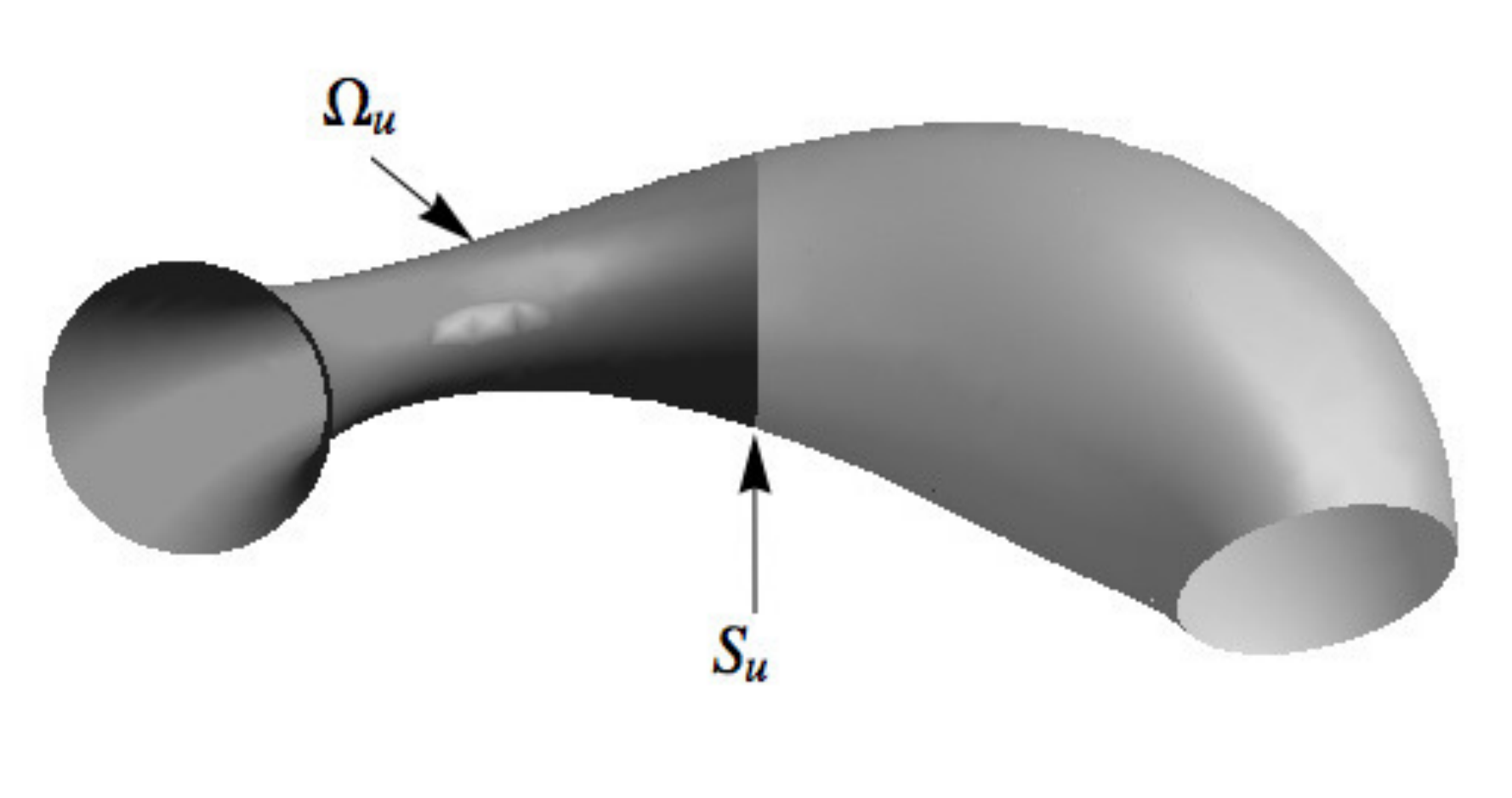}\protect\caption{\label{fig:OmegaAndSAtU}Region $\Omega_{u}$ and cross section $S_{u}$}
\end{figure}

Let us assume that this process is modelled by the continuity equation
\begin{equation}
\der Pt+\hbox{div}(\bm{J})=0,\label{eq:Continuity Equation}
\end{equation}
where $P=P(x,y,z,t)$ is a real valued density function and $\bm{J}=\bm{J}(x,y,z,t)$
is the corresponding flux field. We will apply a dimensionality reduction
technique to this equation as follows. Let $\Omega$ be parametrized
by a smooth map $\bm{\varphi}$ of the form 
\[
\bm{\varphi}(u,v,w)=(x(u,v,w),y(u,v,w),z(u,v,w)),
\]
where $u_{1}\leq u\leq u_{2},v_{1}\leq v\leq v_{2}$ and $w_{1}\leq w\leq w_{2}$.
The parametrization $\bm{\varphi}$ allows us to express $P$ and
$\bm{J}$ in terms of the $u,v,w$ coordinates by letting 
\begin{eqnarray*}
P(u,v,w,t) & = & P(x(u,v,w),y(u,v,w),z(u,v,w),t),\\
\bm{J}(u,v,w,t) & = & \bm{J}(x(u,v,w),y(u,v,w),z(u,v,w),t).
\end{eqnarray*}
For each $u$ we will let $\Omega_{u}$ be the sub-region of $\Omega$
consisting of the points of the form $\bm{\varphi}(s,v,w)$ such that
$u_{1}\leq s\leq u$, and $S_{u}$ be the cross section parametrized
by the map $(v,w)\mapsto\bm{\varphi}(u,v,w)$ (see Figure \ref{fig:OmegaAndSAtU}).

\subsection*{Dimensional reduction of the continuity equation}

From calculus in several variables, the total concentration of $P$
in $\Omega_{u}$ is given by 
\[
C(u,t)=\int_{u_{1}}^{u}\left(\int_{w_{1}}^{w_{2}}\int_{v_{1}}^{v_{2}}P(s,v,w,t)\det(\bm{\varphi}'(s,v,w))dvdw\right)ds,
\]
where $\bm{\varphi}'$ is the Jacobian matrix of $\bm{\varphi}$.
The \emph{effective density} $p$ is defined as
\begin{equation}
p(u,t)=\d Cu(u,t)=\int_{w_{1}}^{w_{2}}\int_{v_{1}}^{v_{2}}P(u,v,w,t)\det(\bm{\varphi}'(u,v,w))dvdw\label{eq:effectiveDensityFormula}
\end{equation}
and the \emph{effective flux} $j$ by 
\[
j(u,t)=\int_{w_{1}}^{w_{2}}\int_{v_{1}}^{v_{2}}\bm{J}(u,v,w,t)\cdot\left(\der{\bm{\varphi}}v(u,v,w)\times\der{\bm{\varphi}}w(u,v,w)\right)dvdw,
\]
where we have denoted the dot product by $\cdot$ and the cross product
by $\times$. The quantity $p(u,t)$ measures the concentration density
at time $t$ along the cross section $S_{u}$, and $j(u,t)$ measures
the flux density along $S_{u}$. Let $\partial\Omega$ denote the
border of the region $\Omega$ and assume that there is no flux of
$P$ across $\partial\Omega-(S_{u_{1}}\cup S_{u_{2}})$. Then by using
the continuity equation (\ref{eq:Continuity Equation}) and the divergence
theorem we obtain the effective continuity equation 
\begin{equation}
\der pt(u,t)+\der ju(u,t)=0.\label{eq:reducedContinuity}
\end{equation}

\subsection*{Diffusion equation}

By imposing Fick's law 
\[
\bm{J}=-D\bm{\nabla}P,
\]
where $\bm{\nabla}P$ denotes the gradient of $P$ in the spatial
directions and $D$ is a constant diffusion coefficient, the continuity
equation (\ref{eq:Continuity Equation}) becomes \emph{the diffusion
equation} 
\[
\der Pt=D\Delta P,
\]
where $\Delta$ is the laplacian operator given by 
\[
\Delta=\frac{\partial^{2}}{\partial x^{2}}+\frac{\partial^{2}}{\partial y^{2}}+\frac{\partial^{2}}{\partial z^{2}}.
\]
In this case, the 1-dimensional effective flux becomes 
\begin{equation}
j(u,t)=-D\int_{w_{1}}^{w_{2}}\int_{v_{1}}^{v_{2}}\bm{\nabla}P(u,v,w,t)\cdot\left(\der{\bm{\varphi}}v(u,v,w)\times\der{\bm{\varphi}}w(u,v,w)\right)dvdw,\label{eq:reducedFluxFormula}
\end{equation}
where 
\[
\bm{\nabla}P(u,v,w,t)=\left(\der Px(\bm{\varphi}(u,v,w),t),\der Py(\bm{\varphi}(u,v,w),t),\der Pz(\bm{\varphi}(u,v,w),t)\right).
\]

\section{A generalized Fick-Jacobs equation on the normal bundle of a 3-dimensional
curve : infinite transversal diffusion rate case }

\label{generalized-fick-jacobs}

In this section we derive a generalized Fick-Jacobs equation and a
new formula for the effective diffusion coefficient (corresponding
to the infinite transversal diffusion rate) for channels that \textquotedbl{}follow\textquotedbl{}
a base curve in 3-dimensional space.

\subsection*{Channel set-up}

Let $\bm{\alpha}=\bm{\alpha}(u)$ be a curve in three dimensional
space parametrized by the arc-length parameter $u$, and consider
scalar functions $\eta=\eta(u,v,w),\beta=\beta(u,v,w)$. Let $\Omega$
be the channel-like region parametrized by the map 
\begin{equation}
\bm{\varphi}(u,v,w)=\bm{\alpha}(u)+\eta(u,v,w)\hat{\bm{N}}(u)+\beta(u,v,w)\hat{\bm{B}}(u),\label{eq:NormalBundleParametrisation}
\end{equation}
where $\hat{\bm{N}}$ and $\hat{\bm{B}}$ are the normal and binormal
fields of $\bm{\alpha}$ (see Appendix 1). In this case, each cross
section $S_{u}$ is contained in the plane passing through $\bm{\alpha}(u)$
and spanned by the vectors $\hat{\bm{N}}(u)$ and $\hat{\bm{B}}(u)$.
By having arbitrary smooth functions $\beta$ and $\eta$ as coefficients
we can generate very general cross sections $S_{u}$. By using the
Frenet-Serret formulae we obtain 
\[
\begin{array}{cccccrccr}
\d{\bm{\varphi}}u & = & (1-\eta\kappa)\hat{\bm{T}} & + & \left(\der{\eta}u-\beta\tau\right) & \hat{\bm{N}} & + & \left(\der{\beta}u+\eta\tau\right) & \hat{\bm{B}},\\
\der{\bm{\varphi}}v & = &  &  & \der{\eta}v & \hat{\bm{N}} & + & \der{\beta}v & \hat{\bm{B}},\\
\der{\bm{\varphi}}w & = &  &  & \der{\eta}w & \hat{\bm{N}} & + & \der{\beta}w & \hat{\bm{B}},
\end{array}
\]
where $\kappa$ and $\tau$ are the curvature and torsion functions
associated to $\bm{\alpha}$. Since $\hat{\bm{T}},\hat{\bm{N}}$ and
$\hat{\bm{B}}$ form an orthonormal basis, the derivative $\bm{\varphi}'$
of $\bm{\varphi}$ can be represented by the following matrix 
\[
[\bm{\varphi}']=\left(\begin{array}{ccc}
1-\eta\kappa & \der{\eta}u-\beta\tau & \der{\beta}u+\eta\tau\\
0 & \der{\eta}v & \der{\beta}v\\
0 & \der{\eta}w & \der{\beta}w
\end{array}\right),
\]
so that 
\begin{eqnarray}
\det(\bm{\varphi}') & = & \omega_{S}(1-\eta\kappa),%
%\der{\bm{\varphi}}v\times\der{\bm{\varphi}}w
\label{eq:phiprime}
\end{eqnarray}
where 
\[
\omega_{S}=\det\left(\begin{array}{cc}
\der{\eta}v & \der{\eta}w\\
\der{\beta}v & \der{\beta}w
\end{array}\right),
\]
and 
\begin{eqnarray}
%\det(\bm{\varphi}')
\der{\bm{\varphi}}v\times\der{\bm{\varphi}}w & = & \omega_{S}\hat{\bm{T}}.\label{eq:phivphiw}
\end{eqnarray}
The map $(v,w)\mapsto\omega_{S}(u,v,w)$ is the area density function
of the cross section $S_{u}$, so that 
\[
A(u)=\int_{w_{1}}^{w_{2}}\int_{v_{1}}^{v_{2}}\omega_{S}(u,v,w)dvdw.
\]
is the area of $S_{u}$. Given a function $f=f(v,w)$, its integral
on $S_{u}$ is given by 
\[
\int_{S_{u}}f=\int_{w_{1}}^{w_{2}}\int_{v_{1}}^{v_{2}}f(v,w)\omega_{S}(u,v,w)dvdw,
\]
i.e. we integrate $f$ over $S_{u}$ by using the area element $\omega_{S}(u,v,w)dvdw$.
The average value of $f$ over $S_{u}$ is then expressed as 
\[
\langle f\rangle_{u}=\frac{1}{A(u)}\int_{S_{u}}f.
\]
In order to simplify notation, we will write $\langle f\rangle$ for
the function $u\mapsto\langle f\rangle_{u},$.

\subsection*{Infinite transversal diffusion rate}

The assumption of \emph{infinite transversal diffusion rate} means
that $P$ is independent of the variables $v$ and $w$. In this case,
we have that the effective density (\ref{eq:effectiveDensityFormula})
is given by 
\begin{equation}
p(u,t)=\omega(u)P(u,t)\label{eq:EffectiveDiffInfiniteTransversalCase}
\end{equation}
where 
\[
\omega(u)=\int_{v_{1}}^{v_{2}}\int_{w_{1}}^{w_{2}}\det(\bm{\varphi}'(u))dvdw.
\]
The function $\omega(u)$ is the volume density function with respect
to $u$, so that 
\[
V(u)=\int_{u_{0}}^{u}\omega(s)ds
\]
is the volume of the region $\Omega_{u}$. By using formula (\ref{eq:phiprime})
we obtain 
\begin{eqnarray*}
\omega(u) & = & \int_{S_{u}}(1-\kappa\eta)\\
 & = & A(u)(1-\kappa(u)\langle\eta\rangle_{u}).
\end{eqnarray*}

To compute the effective flux (\ref{eq:reducedFluxFormula}) observe
that $P$ is constant along the planes passing through $\bm{\alpha}(u)$
and spanned by $\hat{\bm{N}}(u)$ and $\hat{\bm{B}}(u)$. Hence $\bm{\nabla}P$
is orthogonal to $\hat{\bm{N}}$ and $\hat{\bm{B}}$ so that 
\begin{eqnarray*}
\der Pu & = & \bm{\nabla}P\cdot\frac{\partial\bm{\varphi}}{\partial u}\\
 & = & (1-\eta\kappa)\bm{\nabla}P\cdot\hat{\bm{T}},
\end{eqnarray*}
where $\bm{\nabla}P$ is the gradient of $P$ with respect to the
$x,y,z$ variables. Using this and formulas (\ref{eq:reducedFluxFormula})
and (\ref{eq:phivphiw}), we obtain 
\begin{equation}
j(u,t)=-D\der Pu(u,t)\int_{S_{u}}(1-\eta\kappa)^{-1},\label{eq:firstj}
\end{equation}
where $\kappa$ only depends on $u$. By using equation (\ref{eq:EffectiveDiffInfiniteTransversalCase})
and letting 
\begin{eqnarray}
\DD(u) & = & D\left(\frac{\int_{S_{u}}(1-\kappa\eta)^{-1}}{\int_{S_{u}}(1-\kappa\eta)}\right)\label{eq:DFormula}\\
 & = & D\left(\frac{\langle(1-\kappa\eta)^{-1}\rangle_{u}}{1-\kappa\langle\eta\rangle_{u}}\right),\nonumber 
\end{eqnarray}
formula (\ref{eq:firstj}) for $j$ can be written as 
\begin{equation}
j(u,t)=-\DD(u)\omega(u)\dero u\left(\frac{p(u)}{\omega(u)}\right).\label{eq:lastj}
\end{equation}

\subsection*{Generalized Fick-Jacobs equation and effective diffusion coefficient}

If we substitute formula (\ref{eq:lastj}) into the effective continuity
equation (\ref{eq:reducedContinuity}) we obtain the following \emph{generalized
Fick-Jacobs equation} 
\begin{equation}
\der pt(u,t)=\dero u\left(\DD(u)\omega(u)\dero u\left(\frac{p(u,t)}{\omega(u)}\right)\right),\label{eq:FickJacobsEqInCurveBundle}
\end{equation}
which, in turn, casts 
\begin{equation}
\DD(u)=D\left(\frac{\langle(1-\kappa\eta)^{-1}\rangle_{u}}{1-\kappa\langle\eta\rangle_{u}}\right).\label{eq:effective-diffusion-coefficient}
\end{equation}
as the \emph{effective diffusion coefficient}.
\begin{rem*}
When $\kappa=0$, we have that $\DD\equiv D$ and $\omega(u)=A(u)$,
and the above generalized Fick-Jacobs becomes the classical Fick-Jacobs
equation 
\[
\der pt(u,t)=D\dero u\left(A(u)\dero u\left(\frac{p(u,t)}{A(u)}\right)\right).
\]

\end{rem*}

\subsection*{Central curve}

From the definition of $\bm{\varphi}$ we have that 
\[
\langle\bm{\varphi}\rangle_{u}=\bm{\alpha}(u)+\langle\eta\rangle_{u}\hat{\bm{N}}(u)+\langle\beta\rangle_{u}\hat{\bm{B}}(u).
\]
Hence, $\langle\eta\rangle_{u}$ is the $\hat{\bm{N}}(u)$ component
of $\langle\bm{\varphi}\rangle_{u}$ when taking $\bm{\alpha}(u)$
as reference point. We will refer to the curve $\langle\bm{\varphi}\rangle$
as the central curve of the channel defined by $\bm{\varphi}$.
\begin{rem*}
Since the volume of the region $\Omega_{u}$ is given by 
\[
V(u)=\int_{u_{1}}^{u}A(u)(1-\kappa(u)\langle\eta\rangle_{u})du,
\]
when $\bm{\alpha}$ and $\langle\bm{\varphi}\rangle$ coincide, we
have $\langle\eta\rangle\equiv0$ and 
\[
V(u)=\int_{u_{1}}^{u}A(u)du.
\]
For $\bm{\alpha}$ a circle, the last formula is the well known Pappus
theorem which establishes how to compute the volumes of solids of
revolution.
\end{rem*}

\subsection*{Geometric moments}

We can get a better understanding of the function $\langle(1-\eta\kappa)^{-1}\rangle$
appearing in the numerator of $\DD$, by considering the geometric
series expansion 
\[
(1-\eta\kappa)^{-1}=\sum_{n=0}^{\infty}\eta^{n}\kappa^{n}.
\]
Observe that the lower order terms in this series dominate when $\kappa\eta<1$,
i.e. when the $\eta$ coordinates of the channel are far away from
the focal points $\bm{\alpha}+\hat{\bm{N}}/\kappa$ of the base curve
$\alpha$. This last condition is consistent with our narrow channel
assumption. Using the above expansion we can write 
\[
\langle(1-\eta\kappa)^{-1}\rangle=\left(\sum_{i=0}^{\infty}\langle\eta^{i}\rangle\kappa{}^{i}\right),
\]
We will refer to the functions $\langle\eta^{i}\rangle$ as the channel's
$\eta-$moments. Hence, we can write 
\begin{equation}
\DD(u)=\left(\frac{D}{1-\kappa(u)\langle\eta\rangle}\right)\sum_{i=0}^{\infty}\langle\eta^{i}\rangle\kappa^{i}.\label{eq:DSeries}
\end{equation}

\begin{rem*}
Observe that when $\kappa=0$, we have $\DD(u)=D$, and hence all
the geometric information provided by the $\eta$ moments of the channel
is lost. This is, in fact, a good reason why to study the projection
of diffusion along general (non-straight) curves. 
\end{rem*}

\subsection*{The first three geometric moments}

\begin{figure}
\includegraphics[scale=0.55]{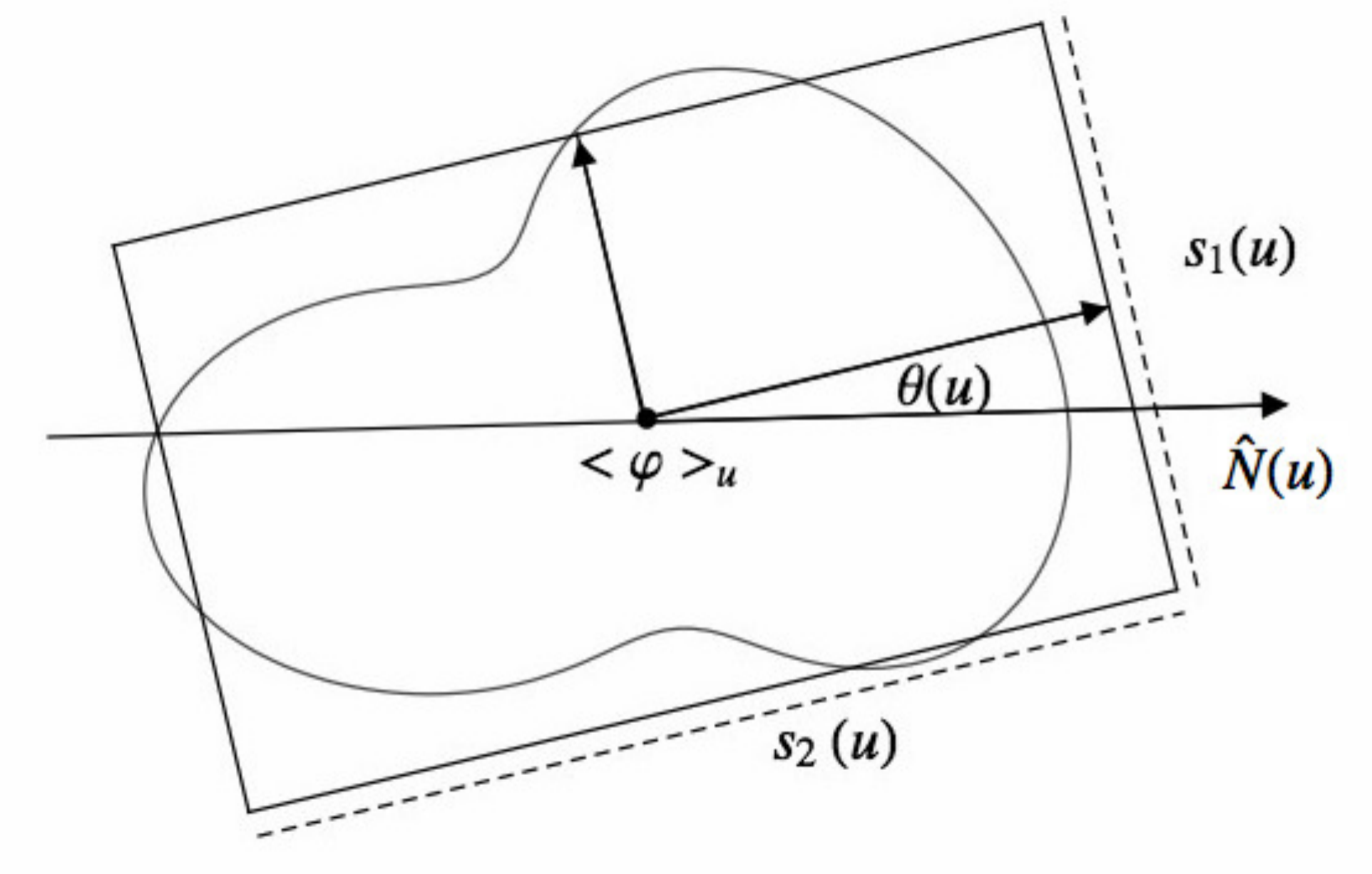}\protect\caption{\label{fig:SecondMoments}Average orientation and sizes of a channel's
cross section with respect to $<\bm{\varphi}>$.}
\end{figure}

We will now use the first three terms in the series (\ref{eq:DSeries})
to relate the effective diffusion coefficient $\DD$ to geometric
properties of the channel. Consider the symmetric matrix 
\[
M(u)=\left(\begin{array}{cc}
a(u) & c(u)\\
c(u) & b(u)
\end{array}\right),
\]
where 
\begin{eqnarray*}
a & = & \langle(\eta-\langle\eta\rangle)^{2}\rangle,\\
b & = & \langle(\beta-\langle\beta\rangle)^{2}\rangle,\\
c & = & \langle(\eta-\langle\eta\rangle)(\beta-\langle\beta\rangle)\rangle.
\end{eqnarray*}
The eigenvectors and eigenvalues of $M(u)$ can be used to measure
the average orientation angle $\theta(u)$ and average sizes $s_{1}(u)$
and $s_{2}(u)$ of the cross section $S_{u}$ in the $\hat{\bm{N}}(u)$
and $\hat{\bm{B}}(u)$ directions with respect to its central point
$\langle\bm{\varphi}\rangle_{u}$ (see Figure \ref{fig:SecondMoments}).
Let $\lambda_{1}(u)$ and $\lambda_{2}(u)$ be the ordered eigenvalues
of $M(u)$ such that $\lambda_{1}(u)\geq\lambda_{2}(u)$. The angle
$\theta(u)$ is the one formed between $\hat{\bm{N}}(u)$ and the
eigenvector of $M(u)$ corresponding to $\lambda_{1}(u)$, and the
functions $s_{1}$ and $s_{2}$ are given by the formulas 
\[
s_{1}(u)=2\sqrt{\lambda_{1}(u)}\quad\and\quad s_{2}(u)=2\sqrt{\lambda_{2}(u)}.
\]
Some simple algebra then shows that 
\begin{eqnarray*}
a & = & \left(\frac{s_{1}}{2}\right)^{2}\cos^{2}(\theta)+\left(\frac{s_{2}}{2}\right)^{2}\sin^{2}(\theta),\\
b & = & \left(\frac{s_{2}}{2}\right)^{2}\cos^{2}(\theta)+\left(\frac{s_{1}}{2}\right)^{2}\sin^{2}(\theta),\\
c & = & \left(\left(\frac{s_{1}}{2}\right)^{2}-\left(\frac{s_{2}}{2}\right)^{2}\right)\cos(\theta)\sin(\theta).
\end{eqnarray*}
Hence 
\[
\langle\eta^{2}\rangle=\langle\eta\rangle^{2}+\left(\frac{s_{1}}{2}\right)^{2}\cos^{2}(\theta)+\left(\frac{s_{2}}{2}\right)^{2}\sin^{2}(\theta).
\]
Using the above formulas and the first three terms of the series (\ref{eq:DSeries})
we obtain the following approximation 
\begin{equation}
\DD\approx\frac{D(1+\langle\eta\rangle\kappa+\left(\langle\eta\rangle^{2}+\left(\frac{s_{1}}{2}\right)^{2}\cos^{2}(\theta)+\left(\frac{s_{2}}{2}\right)^{2}\sin^{2}(\theta)\right)\kappa^{2})}{(1-\kappa\langle\eta\rangle)}.\label{eq:DSecondOrder}
\end{equation}

\subsection*{Higher order moments}

The higher order moments of $\eta$, i.e. the functions $\langle\eta^{i}\rangle$
for $i>2$, contain more subtle information of the geometry of the
channel than that provided by the moments of order $0,1$ and $2$.
For example, in the context of probability distributions concepts
like skewness and kurtosis, which are a measure the asymmetry and
\textquotedbl{}peakedness\textquotedbl{} of a distribution respectively,
involve in their definition moments of order higher than two. These
ideas can be carried onto the case of channels, where we would be
talking about geometric distributions instead of probability distributions.

\section{Applications - Twisted channels with offsets}

\label{applications}

In this section we will apply our results to show how our formula
for the effective diffusion coefficient captures information about
the way the cross section of a channel gyrates with respect to the
Frenet-Serret frame, as well as the effects of offsets from the base
curve. We deduce Ogawa's formula \cite{kn:ogawa} as a particular
case.

We will consider a parametrisation $\bm{\varphi}$ of the form (\ref{eq:NormalBundleParametrisation})
where $\eta$ and $\beta$ are constructed as follows. For a fixed
planar region $R_{0}$ parametrized by the the map 
\begin{equation}
(v,w)\mapsto(\eta_{0}(v,w),\beta_{0}(v,w)),\label{regionR0Param}
\end{equation}
we let $\eta,\beta$ be given by 
\begin{equation}
\left(\begin{array}{c}
\eta(u,v,w)\\
\beta(u,v,w)
\end{array}\right)=\left(\begin{array}{cc}
\cos(\omega u) & -\sin(\omega u)\\
\sin(\omega u) & \cos(\omega u)
\end{array}\right)\left(\begin{array}{c}
\eta_{0}(v,w)\\
\beta_{0}(v,w)
\end{array}\right)+\left(\begin{array}{c}
p(u)\\
q(u)
\end{array}\right)\label{etabeta}
\end{equation}
For a given curve $\bm{\alpha}$, the parametrization $\bm{\varphi}$
with the above $\eta$ and $\beta$, represents a channel constructed
by rotating the region $R_{0}$ with angular velocity $\omega$ (as
we move along the $u$-variable) with respect to the Frenet-Serret
frame of $\bm{\alpha}$, and having offset $p(u)\hat{\bm{N}}(u)+q(u)\hat{\bm{B}}(u)$
from $\bm{\alpha}(u)$.

\subsection{Twisted elliptical cross sections with offsets}

\begin{figure}
\includegraphics[scale=0.35]{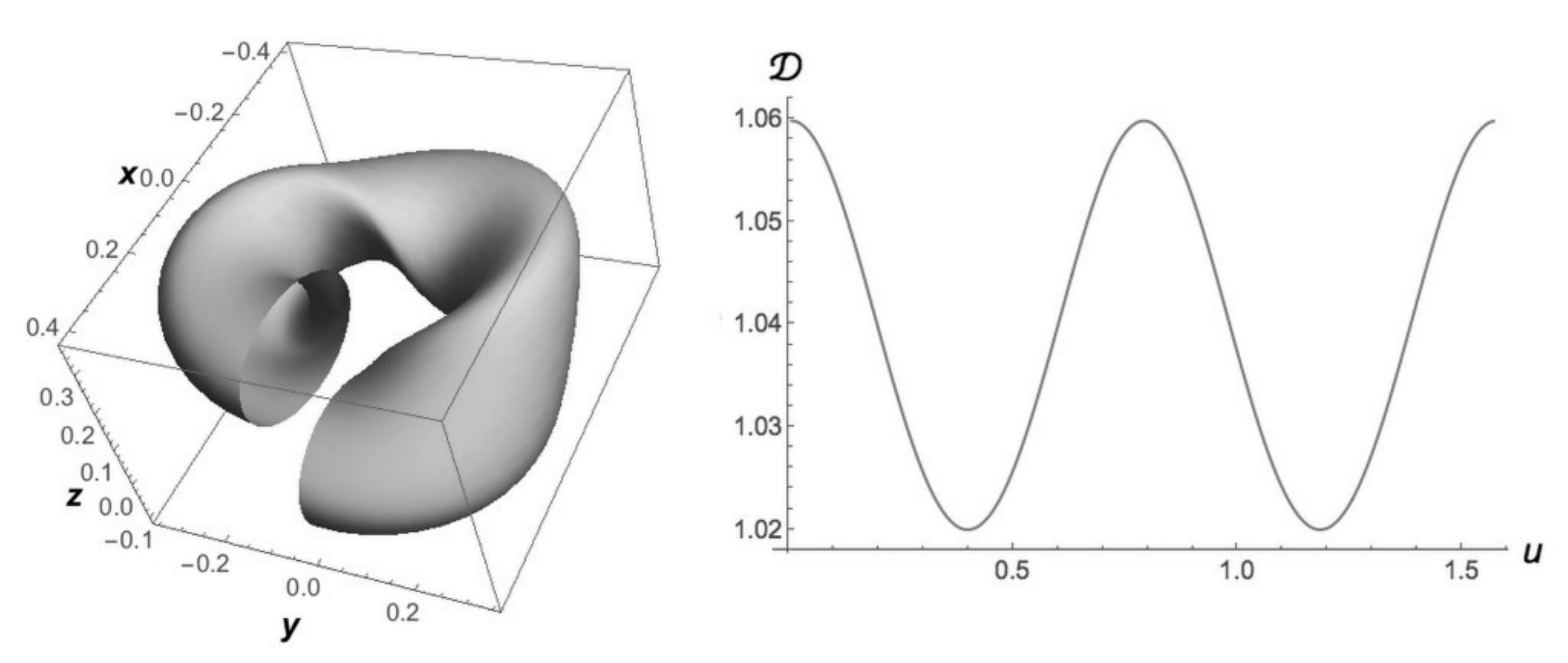} \protect\caption{\label{fig:gyratedEllipse} A twisted elliptical channel over a helix
(left) and the corresponding effective diffusion coefficient function
(right). The relevant parameters are: $D=1,a=1/4,b=1/6,r_{1}=1/6,r_{2}=1/10,p=0,q=0$
and $\omega=4$}
\end{figure}

A solid ellipse with mayor and minor radii $r_{1}$ and $r_{2}$ can
be parametrized by the map (\ref{regionR0Param}) with 
\[
\eta_{0}=vr_{1}\cos(w)\and\beta_{0}=vr_{2}\sin(w),
\]
for $0\leq v\leq1$ and $-\pi\leq w\leq\pi$. If we use $\eta$ and
$\beta$ defined by formula (\ref{etabeta}), then the average sizes
and the area of the channel's cross sections are given by 
\[
s_{1}=r_{1},s_{2}=r_{2}\and A=\pi r_{1}r_{2}.
\]
In this case we can evaluate the integrals in formula (\ref{eq:DFormula})
to obtain (see Appendix 2 for details) 
\begin{equation}
\DD(u)=\frac{2D}{(\kappa(u)R(u))^{2}}\left(1-\frac{\sqrt{(1-p(u)\kappa(u))^{2}-(R(u)\kappa(u))^{2}}}{1-\kappa(u)p(u)}\right),\label{eq:DEllipse}
\end{equation}
where 
\[
R(u)=\sqrt{r_{1}^{2}\cos^{2}(\omega u)+r_{2}^{2}\sin^{2}(\omega u)}.
\]
Observe that if there is no gyration and no offsets of the transversal
cross sections, i.e. $\omega=p=q=0$, then formula (\ref{eq:DEllipse})
for $\DD$ becomes 
\[
\DD(u)=\frac{2D\left(1-\sqrt{1-r_{1}^{2}\kappa^{2}(u)}\right)}{r_{1}^{2}\kappa^{2}(u)}.
\]

\begin{figure}
\includegraphics[scale=0.4]{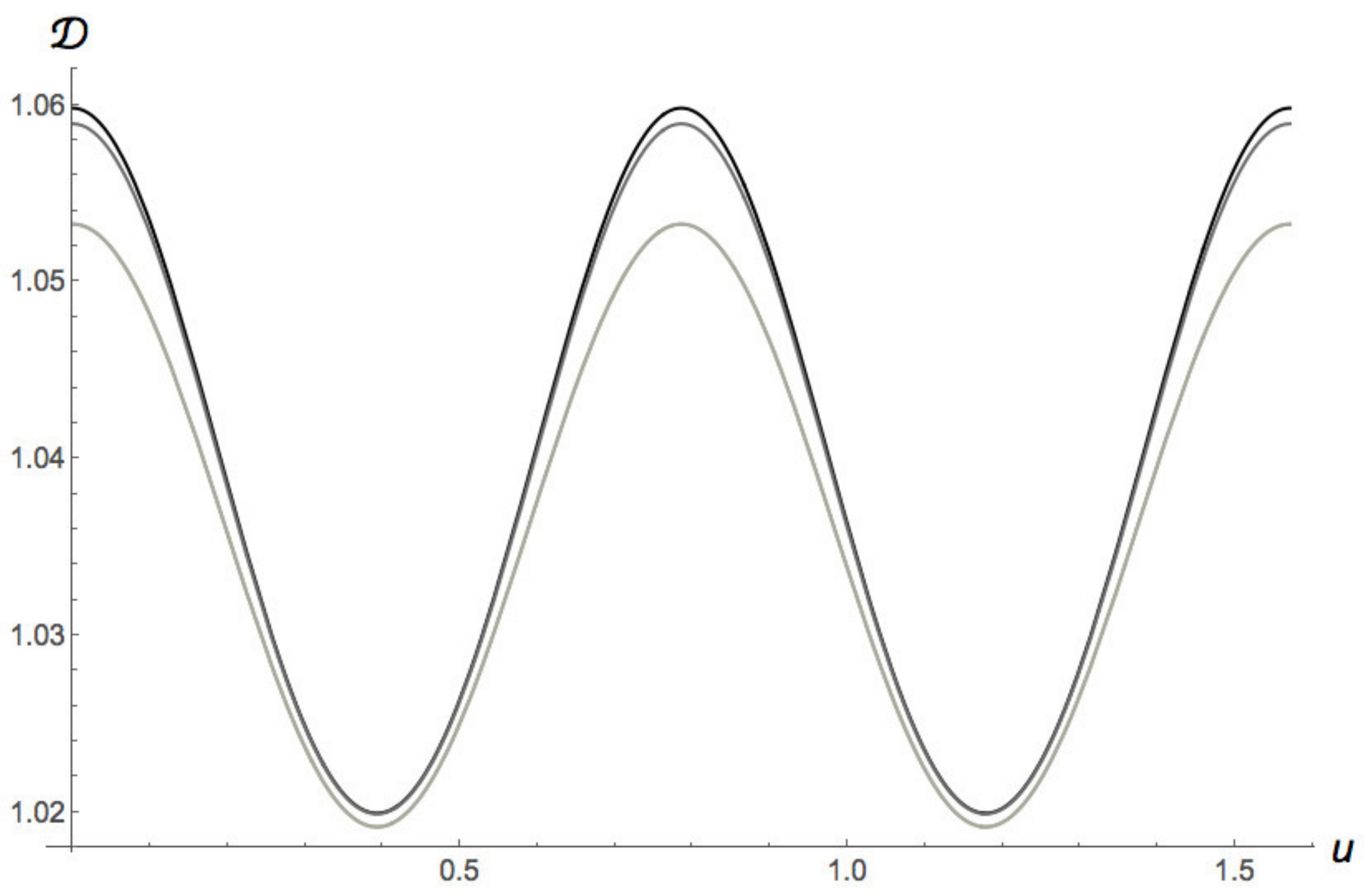} \protect\caption{\label{fig:dSeriesFig} Geometric series approximation to the effective
diffusion coefficient for a channel with gyrating elliptical cross
section. The parameters used to generate these curves are the same
as those in Figure \ref{fig:gyratedEllipse}}
\end{figure}

It is natural to ask how the terms in the series (\ref{eq:DSeries})
approximate our formula (\ref{eq:DEllipse}). Due to the symmetry
of the elliptical sections, the odd terms of this series vanish. Consider
the curves shown in Figure \ref{fig:dSeriesFig} (counting them from
the bottom to the top). The first curve shows the function obtained
by truncating the series after the second term, which corresponds
to formula (\ref{eq:DSecondOrder}). The second curve shows the function
obtained by truncating the series after the fourth term. Finally,
the top curve is the graph of the effective diffusion coefficient
given by (\ref{eq:DEllipse}).

\subsection{Twisted rectangular cross sections with offsets}

\begin{figure}
\includegraphics[scale=0.35]{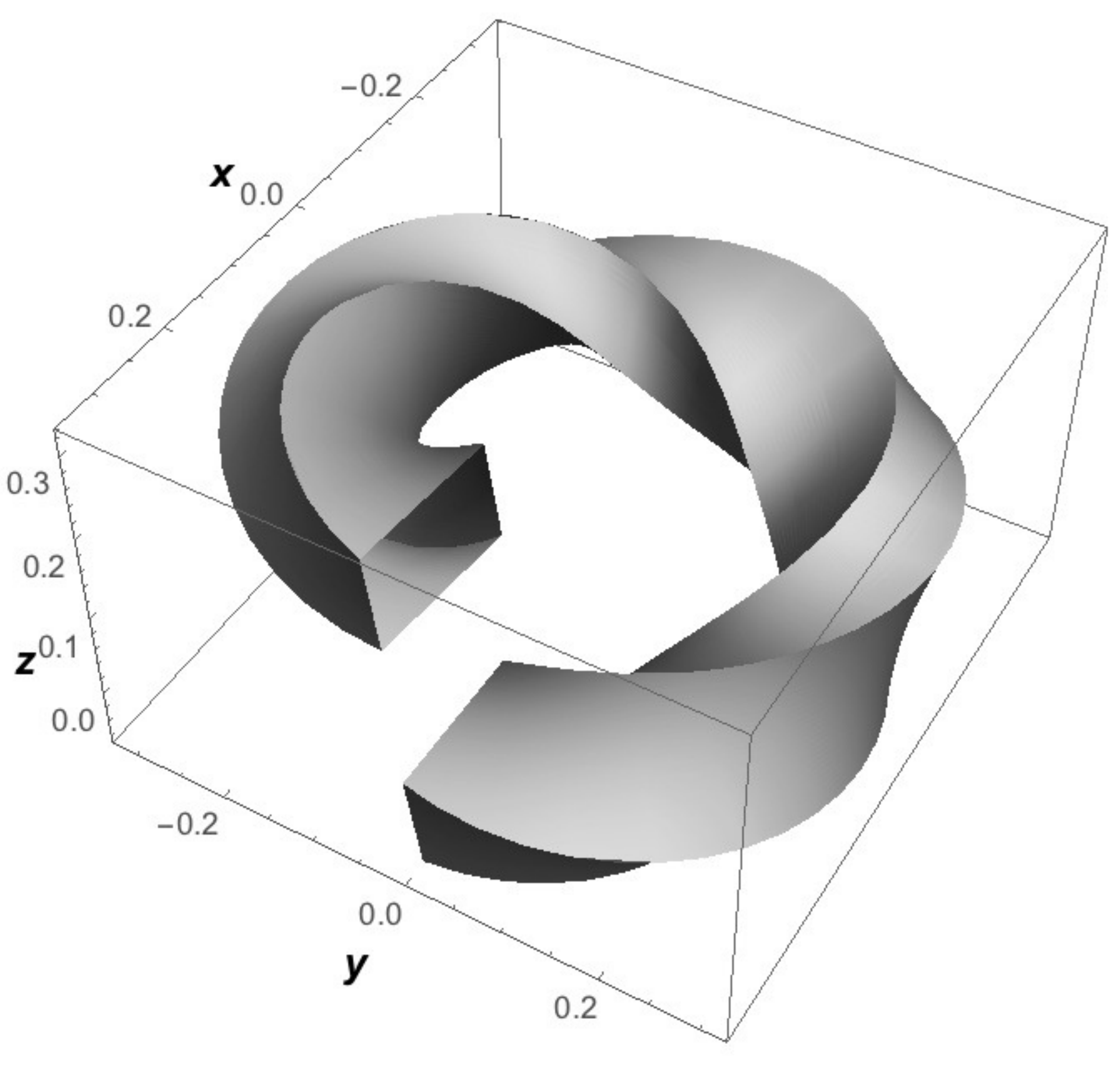}\protect\caption{\label{fig:RectangularChannel} A twisted channel with rectangular
cross section. The parameters used to generate the channel are $a=1/4,b=1/6,d_{1}=1/6,d_{2}=1/10,p=0,q=0$
and $\omega=4$}
\end{figure}

A solid rectangle with sides $d_{1}$ and $d_{2}$ can be parametrized
by a map of type (\ref{regionR0Param}) by letting 
\[
\eta_{0}=v\and\beta_{0}=w,
\]
where $-d_{1}/2\leq v\leq d_{1}/2$ and $-d_{2}/2\leq w\leq d_{2}/2$.
We then have that 
\[
s_{1}=d_{1}/\sqrt{3},\quad s_{2}=d_{2}/\sqrt{3}\and A=d_{1}d_{2}.
\]
In this case we can compute the integrals in formula (\ref{eq:DFormula})
to obtain (see Appendix 2 for details) 
\begin{equation}
\DD(u)=\frac{D\left(\sum_{i=1}^{4}(-1)^{i+1}\gamma_{i}(u)\log(\gamma_{i}(u))\right)}{(d_{1}d_{2}\kappa(u))^{2}(1-\kappa(u)p(u))\cos(\omega u)\sin(\omega u)},\label{eq:DRectangle}
\end{equation}
where 
\[
\gamma_{i}(u)=1-\kappa(u)(p(u)-(\cos(\omega u),\sin(\omega u))\cdot z_{i})
\]
and 
\[
z_{1}=\frac{1}{2}(d_{1},d_{2}),\quad z_{2}=\frac{1}{2}(d_{1},-d_{2}),\quad z_{3}=\frac{1}{2}(-d_{1},-d_{2})\and z_{4}=\frac{1}{2}(-d_{1},d_{2}).
\]
When there is no gyration, i.e. $\omega=0$, we have that 
\[
\DD(u)=\frac{1}{\kappa(u)d_{1}}\log\left(\frac{1+\kappa(u)(d_{1}/2-p(u))}{1-\kappa(u)(d_{1}/2+p(u))}\right).
\]
For $p=0$, this formula is the one obtained by Ogawa in \cite{kn:ogawa}.

\subsection{Twisted cardioidal cross sections with offsets}

\begin{figure}
\includegraphics[scale=0.35]{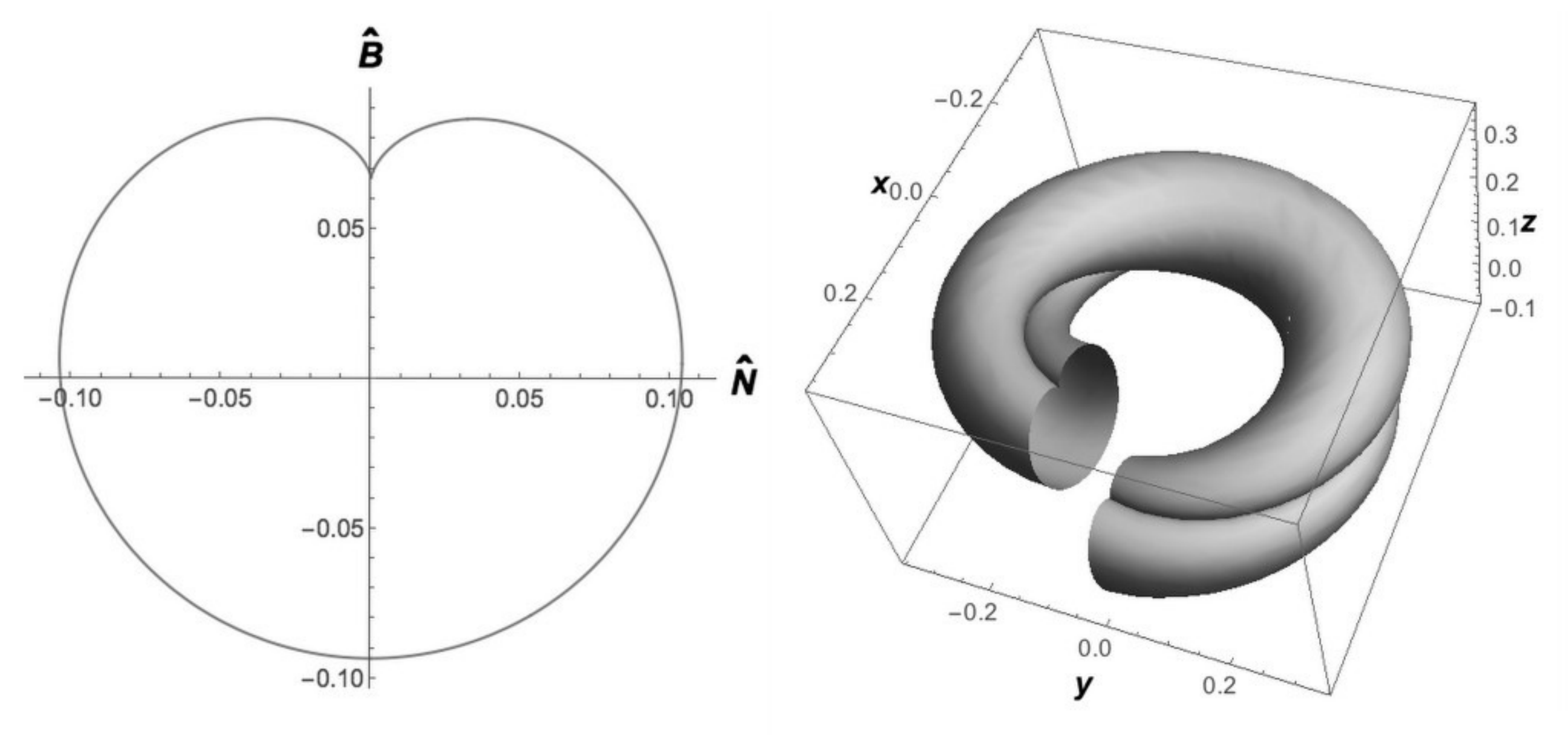}\protect\caption{\label{fig:cardioidChannel} A twisted channel with cardioidal cross
section. The parameters used to generate this channel are $a=1/4,b=1/6,r=1/25,\omega=4,p=0,q=0$.}
\end{figure}

In this case we have that 
\[
\eta_{0}=vr(2\sin(w)-\sin(2w))\and\beta_{0}=vr(2\cos(w)-\cos(2w))+(2/3)r,
\]
where the parameter $r$ is the radius of the circle used to construct
the cardioidal curve. The interior of the region shown in the left
part of Figure \ref{fig:cardioidChannel} is the region parametrized
by the map $(v,w)\mapsto(\eta_{0}(v,w),\beta_{0}(v,w))$ for $0\leq v\leq1$
and $-\pi\leq w\leq\pi$. The right part of the figure shows the channel
resulting from gyrating this cross section over the Frenet-Serret
frame of a helix. Under the above hypotheses we obtain 
\[
s_{1}=\sqrt{7}r,\quad s_{2}=\frac{\sqrt{47}}{3}r\and A=6\pi r^{2}.
\]
In this case, we use the series (\ref{eq:DSeries}) to compute explicit
formulas that approximate $\DD$, and use numerical techniques to
compute the integrals (\ref{eq:DFormula}) in concrete examples.

\subsection{Comparing the elliptical, rectangular and cardioidal cases}

We conclude by comparing the effective diffusion coefficients of the
three types of twisted channels described above. To do a \textquotedbl{}fair\textquotedbl{}
comparison we need to set the parameters of the cross sections so
that their geometries are similar to second (geometric) order. We
do this by equating their width functions $s_{1}$ and $s_{2}$ and
their angle function $\theta$. For a fixed value of the parameter
$r$ of the cardioid, the mayor and minor radii $r_{1}$ and $r_{2}$
of the elliptical cross section must be set to 
\begin{equation}
r_{1}=\sqrt{7}r\and r_{2}=\frac{\sqrt{47}}{3}r,\label{eq:rsfromr}
\end{equation}
and the sides $d_{1}$ and $d_{2}$ of the rectangular cross section
must be set to 
\begin{equation}
d_{1}=\sqrt{21}r\and d_{2}=\sqrt{\frac{47}{3}}r.\label{eq:dsfromr}
\end{equation}
For the angle functions $\theta$ to be equal we simply need to use
the same $\omega$ as the gyrating velocity in all cases.

\begin{figure}
\includegraphics[scale=0.4]{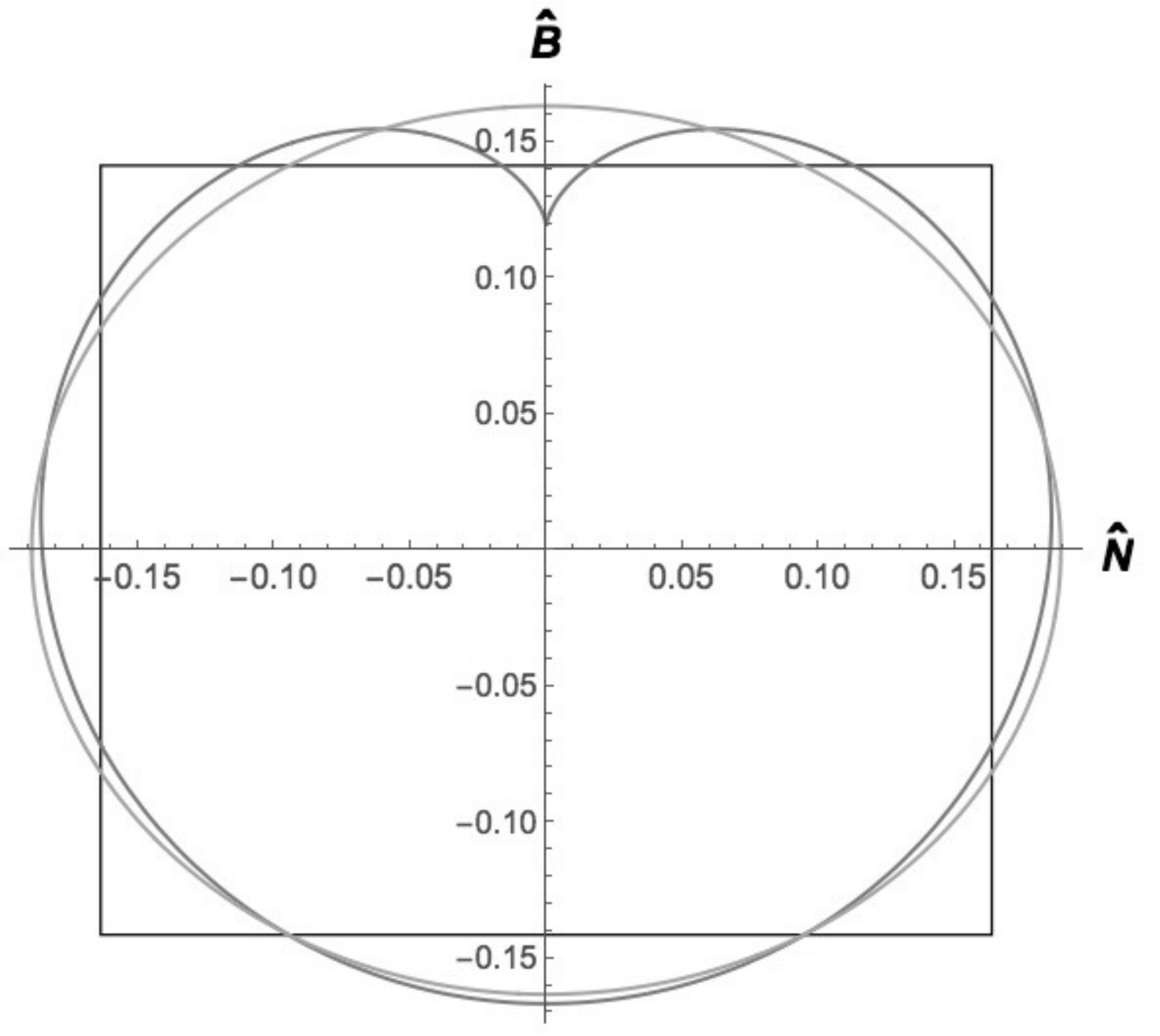}\protect\caption{\label{fig:SectionComparison} Elliptical, rectangular and cardioidal
sections having the same width values $s_{1}$ and $s_{2}$.}
\end{figure}

\begin{figure}
\includegraphics[scale=0.35]{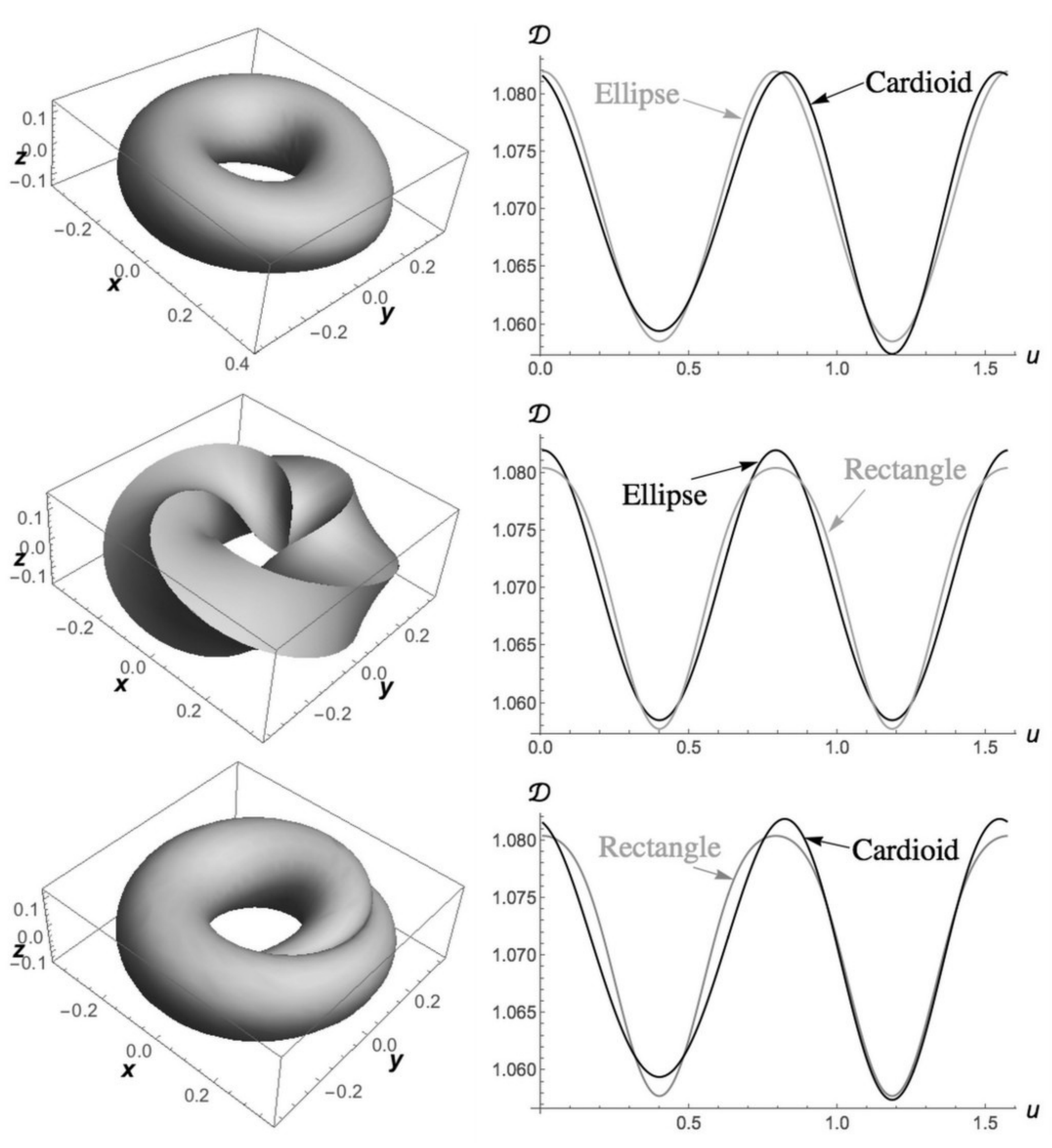}\protect\caption{\label{fig:comp20} Pairwise comparisons of the effective diffusion
coefficients for the twisted elliptical, rectangular and cardioidal
channels with $r=1/20$.}
\end{figure}

\begin{figure}
\includegraphics[scale=0.35]{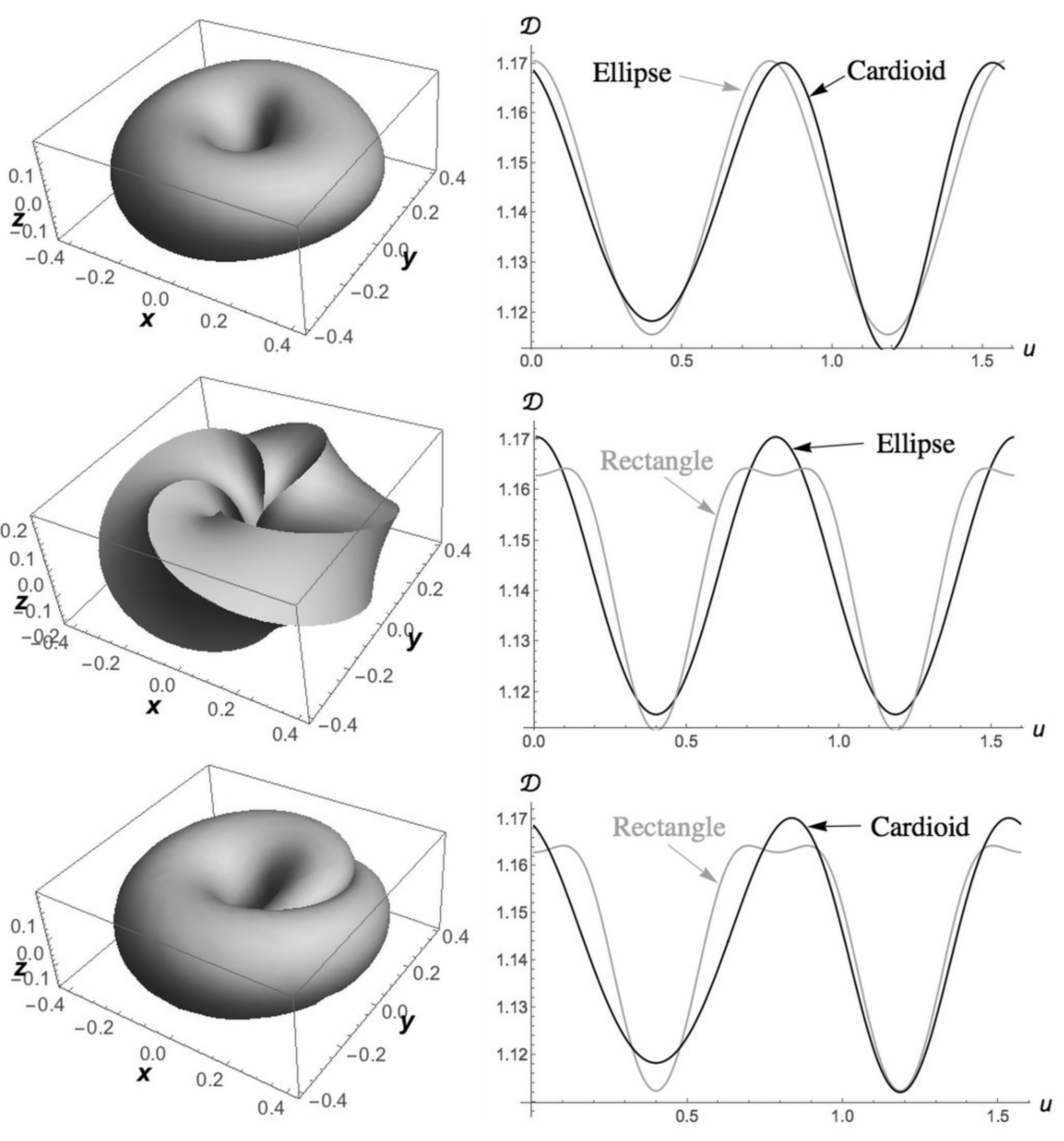}\protect\caption{\label{fig:comp15} Pairwise comparisons of the effective diffusion
coefficients for the twisted elliptical, rectangular and cardioidal
channels with $r=1/15$.}
\end{figure}

We will illustrate the behaviour of the effective diffusion coefficients
in these cases by letting 
\[
a=1/4,b=0,\omega=4.
\]
In Figure \ref{fig:comp20} we show the results obtained from the
above selection of parameters by letting $r=1/20$, and in Figure
\ref{fig:comp15} the results obtained by letting $r=1/15$. The effective
diffusion formulas used in these examples are (\ref{eq:DEllipse})
and (\ref{eq:DRectangle}) for the elliptical and the rectangular
case, and the cardioidal case was computed using numerical integration.
To give an explanation of the behaviour just illustrated, we need
the following description of the focal line.

\subsection*{The focal line and the effective diffusion coefficient}

The focal set of a 3-dimensional curve $\bm{\alpha}$ consists of
the points of the form $\bm{\alpha}(u)+(1/\kappa(u))\hat{\bm{N}}(u)$.
The focal line through such a point is the one having direction $\hat{\bm{B}}(u)$.
The curve $\bm{\alpha}$ used in the examples in Figures \ref{fig:comp20}
and \ref{fig:comp15} is a circle in the $xy$-plane with radius $a$,
and in this case the focal set consists of the origin $(0,0,0)$ and
the corresponding focal lines have direction $\hat{\bm{B}}=(0,0,1)$.
For a given point $\bm{p}=\text{\ensuremath{\bm{\alpha}}}(u)+\eta(u,v,w)\hat{\bm{N}}(u)+\beta(u,v,w)\hat{\bm{B}}(u)$
in a cross section $S_{u}$, the distance of $\bm{p}$ to the corresponding
focal line is 
\[
d_{f}(\bm{p})=\frac{1-\kappa(u)\eta(v,w)}{\kappa(u)}
\]
In order to simplify the arguments, let us assume that $<\eta>=0$
(which holds in our examples). Formula (\ref{eq:DFormula}) can then
be written as 
\begin{equation}
\DD(u)=\frac{1}{A(u)\kappa(u)}\int_{S_{u}}\frac{1}{d_{f}}.\label{eq:focalDiff}
\end{equation}
Thus, we can make the following observations about the examples in
Figures \ref{fig:comp20} and \ref{fig:comp15}. The further away
the cross section $S_{u}$ is from the focal line, the less influence
the geometry (geometric moments information) of the cross section
has on $\DD(u)$. Notice that in Figure \ref{fig:comp20} the effective
diffusion coefficients look more similar to each other than in Figure
\ref{fig:comp15}, where the effective diffusion coefficient for the
rectangular channel has developed extra \textquotedbl{}bumps\textquotedbl{}
due to the proximity of the channel to the focal line. As $r$ tends
to $0$, the cross sections are further away from the focal line,
and the effective diffusion coefficients look more similar to each
other. In the limit when $r=0$, all the effective diffusion coefficients
become equal to $D=1$.

\subsection*{Symmetries}

We will now explain some of the features of the effective diffusion
coefficients shown in Figures \ref{fig:comp20} and \ref{fig:comp15}
in terms of the symmetries of the cross sections with respect to the
normal field $\hat{\bm{N}}$. Observe that we have $0\leq u\leq\pi/2$
and $\omega=4$, so that the angle $\theta=\omega u$ covers a full
cycle from $0$ to $2\pi$.

For $\theta=0,\pi/2,\pi,3\pi/2$ the ellipse is invariant under reflections
with axis given by the normal vector $\hat{\bm{N}}$ at these points.
This explains the four critical points of $\DD$ at $u=0,\pi/8,\pi/4,3\pi/8$.
When $u=0,\pi/4$ the mayor axis of the ellipse faces the focal line
and the corresponding critical points are local maxima. When $u=\pi/8,3\pi/8$
the minor axis of the ellipse faces the focal line and the corresponding
critical points are local minima. This is consistent with our observation
that the closer the cross section is to the focal line, the larger
the effect it has on the effective diffusion coefficient.

For $\theta=\pi/2,3\pi/2$ the cardioid is invariant under reflections
with axis given by the normal vector $\hat{\bm{N}}$ at these points.
This explains the two critical points of $\DD$ at $u=\pi/8,3\pi/8$.
These two points are local minima of $\DD$, since the smallest axis
of the cardioid faces the focal line for these angles. The first local
minimum is smaller than the second because in the first case the \textquotedbl{}dent\textquotedbl{}
of the cardioid is directed towards the focal line and in the second
case this \textquotedbl{}dent\textquotedbl{} faces away from the focal
line. The local maxima of $\DD$ appearing near $u=0$ and $u=\pi/4$
can be explained again by the fact the largest axis of the cardioid
faces the focal line at these angles, and the asymmetry of the the
cardioid with respect to reflection along the normal line explains
the fact that the $u$ values at which this maxima occur, appear with
offsets (to the left and right) to the exact values $0$ and $\pi/4$.

The behaviour of the critical points in the rectangle case can be
explained in a similar way as in the previous two cases, with the
added effect (if the rectangle is close enough to the focal line)
that the corners of the rectangle generate the \textquotedbl{}bumps\textquotedbl{}
on the effective diffusion function (shown in Figure \ref{fig:comp15})
as they get closer to the focal line.

\section{Conclusions and future work}

\label{conclusions}

We have deduced a new formula for the effective diffusion coefficient
$\DD$ of a generalized Fick-Jacobs equation for narrow 3-dimensional
channels. We derived such a formula by projecting the diffusion equation
along the normal directions of a base curve of a narrow channel in
3-dimensional space under the assumption of infinite transversal diffusion
rate, and using tools of differential geometry of curves. Our formula
establishes an explicit relation between some of the channel's geometric
properties (i.e. curvature of the base curve and the geometric moments
of the transversal cross sections) and the corresponding effective
diffusion coefficient. We have also showed that previous estimates
\cite{kn:ogawa} for $\DD$ can be recovered from our formula as particular
cases, and how our formula captures information about the way the
cross sections gyrate with respect to the Frenet-Serret frame.

In future work, we will deal with finite transversal diffusion rate
case. We expect that in that case both tangential and curvature information
of the channel's surface will enter into the formula of the effective
diffusion coefficient.

\section{Appendix 1 - The Frenet-Serret formulas for 3D curves.}

In this appendix we review some basic concepts of the differential
geometry of curves in three dimensional space. The material is standard
and can be found in books such as \cite{kn:spivak2,kn:docarmo}. Consider
a smooth curve in three dimensional space of the form $\bm{\alpha}(s)=(x(s),y(s),z(s))$.
The curve is said to have arc-length parametrization if for all $s$
in the interval $[s_{1},s_{2}]$ we have that 
\[
\left|\d{\bm{\alpha}}s\right|=1\where\left|\d{\bm{\alpha}}s\right|=\sqrt{\left(\d xs\right)^{2}+\left(\d ys\right)^{2}+\left(\d zs\right)^{2}}.
\]
If the above condition holds, then the length of the curve segment
$\bm{\alpha}([s_{1},s])$ is given by 
\[
\hbox{length}(\bm{\alpha}([s_{1},s])=\int_{s_{1}}^{s}\left|\d{\bm{\alpha}}s(a)\right|da=s.
\]
We can construct three orthonormal fields to $\alpha$ given by 
\[
\hat{\bm{T}}=\d{\bm{\alpha}}s,\hat{\bm{N}}=\d{\hat{\bm{T}}}s/\left\Vert \d{\hat{\bm{T}}}s\right\Vert \and\hat{\bm{B}}=\hat{\bm{T}}\times\hat{\bm{N}},
\]
which are know as the tangent, normal and bi-normal fields, respectively.
The orthonormality conditions on these fields imply the existence
of scalar functions $\kappa=\kappa(u)$ and $\tau=\tau(u)$, known
as the curvature and torsion, such that 
\begin{eqnarray*}
\d{\hat{\bm{T}}}s & = & \kappa\hat{\bm{N}},\\
\d{\hat{\bm{N}}}s & = & -\kappa\hat{\bm{T}}+\tau\hat{\bm{B}},\\
\d{\hat{\bm{B}}}s & = & -\tau\hat{\bm{N}}.
\end{eqnarray*}
These formulas are known in the literature as the Frenet-Serret formulas,
and the fields $\hat{\bm{T}},\hat{\bm{N}},\hat{\bm{B}}$ as the Frenet-Serret
frame. The curvature function measures the deviation of $\bm{\alpha}$
of being a straight line, and $\tau$ the deviation of $\bm{\alpha}$
 from being in a plane.

As an example, consider a helix of radius $a>0$ and pitch $b>0$
parametrized by 
\[
s\mapsto(a\cos(s),a\sin(s),bs).
\]
The arc-length parametrisation of this curve is 
\[
\alpha(u)=\left(a\cos\left(\sqrt{1-b^{2}}u/a\right),a\sin\left(\sqrt{1-b^{2}}u/a\right),bu\right),
\]
and the corresponding curvature and torsion of this curve are 
\[
\kappa=\frac{a}{a^{2}+b^{2}}\and\tau=\frac{b}{a^{2}+b^{2}}.
\]

\section{Appendix 2 - Details on the computation of the effective diffusion
coefficient}

In our computation of the effective diffusion coefficient we have
used the formula (\ref{eq:DFormula}), which can be written (when
$<\eta>=p$) explicitly as 
\begin{equation}
\DD(u)=\left(\frac{D}{A(u)(1-\kappa(u)p(u))}\right)\int_{v_{1}}^{v_{2}}\int_{w_{1}}^{w_{2}}\left(\frac{\omega_{S}(u,v,w)}{1-\kappa(u)\eta(u,v,w)}\right)dvdw,\label{eq:FullD}
\end{equation}
where 
\[
\omega_{S}(u,v,w)=\det\left(\begin{array}{cc}
\der{\eta}v(u,v,w) & \der{\eta}w(u,v,w)\\
\der{\beta}v(u,v,w) & \der{\beta}w(u,v,w)
\end{array}\right).
\]
For completeness, we will now expand some details regarding the computations
of formulas (\ref{eq:DEllipse}) and (\ref{eq:DRectangle}) from (\ref{eq:FullD}).
We calculate the above integral by using Fubini's Theorem. We do this
by finding a function $H=H(u,v,w)$ such that 
\begin{equation}
\frac{\partial H}{\partial v\partial w}(u,v,w)=\Omega(u,v,w),\label{eq:antiderivative}
\end{equation}
where 
\begin{equation}
\Omega(u,v,w)=\frac{\omega_{S}(u,v,w)}{1-\kappa(u)\eta(u,v,w)}.\label{eq:omegaD}
\end{equation}
We then have 
\begin{equation}
\DD(u)=\left(\frac{D}{A(u)(1-\kappa(u)p(u))}\right)\sum_{i,j=1}^{2}(-1)^{i+j}H(v_{i},w_{j}).\label{eq:sumD}
\end{equation}

\subsection*{Elliptical case}

The integrand function is 
\[
\Omega(u,v,w)=\frac{r_{1}r_{2}v}{1-\kappa(u)\left(p(u)-r_{2}v\sin(w)\sin(\omega u)+r_{1}v\cos(w)\cos(\omega u)\right)}.
\]
From equation (\ref{eq:antiderivative}) we have that 
\[
\der Hw=\frac{r_{1}r_{2}(\kappa(p-\eta)-(1-\kappa p)\log(2(1-\kappa\eta)))}{\kappa^{2}(p-\eta_{1})^{2}},
\]
where $\eta_{1}(u,w)=\eta(u,1,w)$. We then obtain 
\begin{equation}
H=\frac{r_{1}r_{2}}{\kappa^{2}R}\left((1-\kappa p)\left(\frac{S\log(2(1-\kappa\eta))}{p-\eta_{1}}+w\right)-2Q\hbox{arctanh}(T)\right)\label{eq:ePrim}
\end{equation}
where 
\begin{eqnarray*}
R(u) & = & r_{1}^{2}\cos^{2}(\omega u)+r_{2}^{2}\sin^{2}(\omega u)\\
Q(u,v) & = & \sqrt{v^{2}\kappa^{2}(u)R(u)-(1-p(u)\kappa(u))^{2}}\\
T(u,v,w) & = & \frac{(1-\kappa(u)(p(u)-r_{1}v\cos(\omega u))\tan\left(\frac{w}{2}\right)+\kappa(u)r_{2}v\sin(\omega u)}{Q(u,v)}\\
S(u,w) & = & r_{1}\sin(w)\cos(\omega u)+r_{2}\cos(w)\sin(\omega u)\\
\end{eqnarray*}
By using formulas (\ref{eq:sumD}) and (\ref{eq:ePrim}) we obtain
(\ref{eq:DEllipse}).

When trying to directly evaluate the quantities $H(v_{i},w_{j})$
in formula (\ref{eq:sumD}) it turns out that they are not well defined
for the values $w=-\pi,\pi$. We solve this problem by letting 
\[
H(v_{i},\pi)=\lim_{w\mapsto\pi^{-}}H(v_{i},w)\and H(v_{i},-\pi)=\lim_{w\mapsto-\pi^{+}}H(v_{i},w)
\]

\subsection*{Rectangular case}

The integrand function is 
\[
\Omega(u,v,w)=\frac{d_{1}d_{2}}{4\left(1-\kappa(u)\left(p(u)+\frac{1}{2}d_{1}v\cos(\omega u)-\frac{1}{2}d_{2}w\sin(\omega u)\right)\right)}.
\]
From equation (\ref{eq:antiderivative}) we have that 
\[
\der Hw=-\frac{d_{2}\log(2(1-\kappa\eta))}{2\kappa\cos(\omega u)}
\]
and 
\begin{equation}
H=\frac{\log(2(1-\eta\kappa))(d_{1}\kappa v\cot(\omega u)-d_{2}\kappa w+2(\gamma\kappa-1)\csc(\omega u))+d_{2}\kappa w}{2\cos(\omega u)\kappa^{2}}.\label{eq:rPrim}
\end{equation}
By using formulas (\ref{eq:rPrim}) and (\ref{eq:sumD}) we obtain
(\ref{eq:DRectangle}).

\section{Acknowledgments}

The second author would like to thank the International Centre for
Theoretical Physics (ICTP, Italy) and the Institut des Hautes Études
Scientifiques (IHÉS, France) for their hospitality and support.

\bibliographystyle{unsrt}
\bibliography{myBib}

\end{document}